# Development of a Web-based Research Consortium Database Management System: Advancing Data-driven and Knowledge-based Project Management

Mitch Arkeen Salvador

Central Luzon State University, Science City of Muñoz, Philippines, salvador.arkeen@clsu.edu.ph

Khavee Agustus Botangen[*]

Central Luzon State University, Science City of Muñoz, Philippines, kwbotangen@clsu.edu.ph

Mary Camille Rabang

Central Luzon State University, Science City of Muñoz, Philippines, rabang.mary@clsu2.edu.ph

Ivan Christian Salinas

Central Luzon State University, Science City of Muñoz, Philippines, banbansalinas@clsu.edu.ph

Marlon Naagas

Central Luzon State University, Science City of Muñoz, Philippines, manaagas@clsu.edu.ph

Angelika Balagot

Central Luzon State University, Science City of Muñoz, Philippines, angelikabalagot@clsu.edu.ph

The Central Luzon Agriculture, Aquatic and Natural Resources Research and Development Consortium (CLAARRDEC), comprising 29 member institutions, faces challenges in effectively monitoring and evaluating their R&D activities. To address these challenges, they seek to harness digital technology for data management and real-time monitoring. This paper presents the development of a web-based database and real-time monitoring system aimed at enhancing data collection, storage, retrieval, and utilization within the consortium. The system consists of two key components: *i*) a data management module, designed to facilitate project data collection from member institutions, and *ii*) a real-time monitoring module for report generation and analytics at the CLAARRDEC main office. Successful deployment of the system not only fosters information sharing, collaboration, and informed decision-making but also empowers member institutions to monitor their own R&D engagements. Furthermore, the system's potential extends beyond CLAARRDEC, as it could be utilized by other research consortia in the Philippines.

---

[*] Corresponding author

# 1 INTRODUCTION

The Central Luzon Agriculture, Aquatic, and Natural Resources Research and Development Consortium (CLAARRDEC) comprises 29 research and development (R&D) institutions, including state universities and colleges within the Central Luzon Region of the Philippines. These institutions, referred to as consortium member institutions (CMIs), are responsible for implementing numerous government-funded R&D projects. CLAARRDEC's mandate is to monitor and evaluate these R&D activities in the region, specifically those executed by its member institutions.

Traditionally, oversight of these activities involved site visits, documentation, reporting during meetings, and coordination sessions. Currently, CLAARRDEC's collaborative regional program extends to encompass various aspects of R&D, including management, coordination, utilization of results, capability building, governance, and policy development. Each CMI plays a crucial role in providing essential data through reports and documentation to fulfill the consortium's monitoring goals. However, challenges arise in consolidating data, as reports from different institutions are submitted via email or in hardcopy, primarily serving as supporting documents for the annual report submitted at the end of each year. This difficulty in consolidation hinders the transformation of reports into actionable information, with R&D data dispersed across computers in various institutions. This dispersion poses obstacles to retrieving consolidated information, thereby hindering effective data sharing, program planning, and decision-making. Compounding this challenge is the insufficient manpower at CLAARRDEC's main office to handle the volume of submitted data, further impeding the report consolidation process. Retrieval of relevant information on the consortium's R&D activities is difficult. To address these challenges, there is a pressing need for a streamlined and centralized data management system that facilitates efficient data sharing, enhances program planning, and supports effective decision-making within the consortium.

In this paper we present our work in developing a web-based database management system aimed at addressing the aforementioned concerns. The system consists of two main components. The first is the data acquisition module responsible for gathering, receiving, and processing data from the various CMIs. The second is the report and analytics module tasked with consolidating, analyzing, and presenting data through reports and visualizations. Furthermore, the system facilitates real-time monitoring of R&D activities on two levels. On one hand, a CMI's R&D manager can track the details and statuses of projects they are overseeing. On the other hand, the CLAARRDEC main office can monitor all the member CMIs' project details and statuses, whether individually or for the entire consortium. The system also enables the automatic generation of consolidated reports for each member CMI or for the entire consortium. The success of this work opens the potential integration of knowledge management in the consortium's R&D system. Its future adoption in other regional consortia could contribute to a data-driven monitoring, evaluation, and planning of R&D activities on a national scale.

# 2 LITERATURE REVIEW

Organizations worldwide are increasingly relying on software applications to facilitate their project management processes. Although some still depend on spreadsheet applications and basic electronic calendars, the majority have made investments in more robust applications to streamline project management. These comprehensive applications are commonly referred to as Project Management Information Systems (PMIS) [5, 11]. While each PMIS implementation differs in terms of scope, design, and features, software applications are universally considered an indispensable component of these systems. The effective utilization of a PMIS is contingent upon how critical information is made accessible to all stakeholders and the degree of process automation employed for gathering, integrating data, and disseminating the outputs of data analysis [5].

Numerous PMISs have been developed, catering to both generic use and industry-specific needs as presented in the survey paper by [9]. In the competitive landscape of project management software, various companies vie for market dominance [5]. Notable commercial PMIS options include Microsoft Project Server, Oracle Primavera P6, Autodesk, RIB Software, Huddle, Rally, Asana, Wrike, and TechExcel DevSuite [5, 9]. While some software companies offer single applications capable of managing different project aspects, others provide suites of specialized applications that complement each other and can be integrated with enterprise resource planning implementations. However, the adoption of certain commercial applications is often constrained by considerations of pricing and operational costs. This constraint was evident in a study involving 100 project managers as respondents [5], where concerns were raised about the efficacy of deploying commercial software applications for PMIS implementation. Furthermore, it has been observed that several commercially available software tools are not suitable for certain project management and monitoring practices [9]. Consequently, there has been a plethora of studies aiming to develop useful tools and techniques specifically tailored for the conduct of project management processes within certain domains or industries. These endeavors aim to support project managers in planning, organizing, and controlling projects of varying complexity [9, 1, 13].

In [4] a web-based project management system is proposed for agricultural scientific research in China, integrating modules for system and user administration, financial management, document management, communication management, information dissemination, and progress management. Similarly, [15] introduced a project management system designed for the management and monitoring of projects within a university. The system's functionalities encompass proposal submission, review, progress and terminal report submission, and the display of project results. Additionally, [7] proposed a web-based project management system to oversee and monitor the progress of undergraduate projects, covering stages such as proposal, assessment, approval, submission of phase reports, final reports, presentation scheduling, and the assignment of final grades. The work in [12] introduced an energy data consolidation platform designed to allow users to import energy data and production data from various sources, consolidating it into a comprehensive single dataset. The data consolidation platform is integrated into a web-based energy management system, utilized for generating a variety of reports including demand-side management, tax incentives, mandatory provision of energy data, and customized performance reports.

The establishment of a provincial health data center in South Africa is outlined in [3], aiming to integrate and process individual healthcare data from various sources. This initiative provides the capability to generate information for improved healthcare services, including patient-specific email alerts, a web-based consolidated patient clinical viewing platform, management reports and dashboards, and data releases in response to operational and research data requests. In a similar health data context, [14] proposed a web-based electronic health record (EHR) system for data acquisition and consolidation, simplifying the process of collecting, presenting, and analyzing patient information while adhering to openEHR standards. Additionally, [10] introduced a web-based data management system for a consortium of 24 institutions, capturing radiation therapy dosimetry data along with patient- and physician-reported outcomes and clinical data for a radiation therapy collaborative quality initiative.

The literature consistently emphasizes that PMISs share a common focus on addressing concerns related to the data consolidation process. Recognizing the critical role of PMISs in efficiently consolidating data from diverse sources, this step serves as the foundation for generating relevant reports and insights essential for decision support, strategic planning, and other managerial functions.

## 3 METHODOLOGY

In our system development approach, we embraced a hybrid model combining elements of the Waterfall Model and Agile Model to guide the primary phases of our development process, as illustrated in Figure 1a. The initial stages, such as requirements analysis and design, adhered to the principles of the Waterfall Model, yielding high-level system designs.

Subsequent development (coding) and testing phases were executed using Agile methodology, incorporating a series of iterations. Each iteration commenced with a meeting involving either prospective users or system testers, during which activities like prototype presentations or system component testing took place. Based on the outcomes of these activities, we updated the items listed in the system backlog—a document cataloging user feedback, features, enhancements, bug fixes, and other work items representing system requirements. Backlog refinement sessions were conducted to finalize and elaborate on the items for better comprehension. Each backlog item was assigned an estimated time for implementation by the development team. We then proceeded to implement and integrate these requirements into the existing system, followed by a meeting with users for system demos and testing. This cycle of activities within the development and testing phases was repeated as needed for further refinement and enhancement.

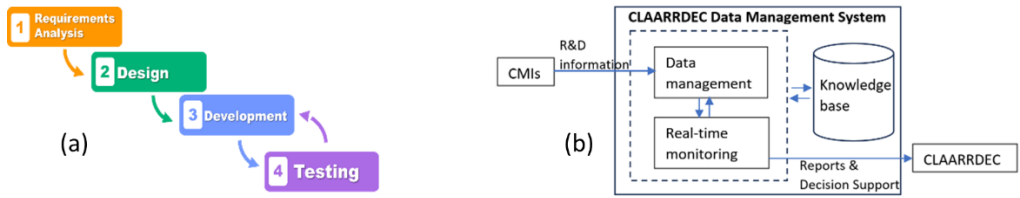

Figure 1: Conceptual diagram of (a) a hybrid of the Waterfall Model and Agile Model, and (b) architectural design of the system.

## 3.1 Requirements Analysis and Design

We engaged in numerous meetings with the management and staff at the CLAARDEC main office to gain insights into the current state of their project management processes, as well as to understand their data needs and capabilities. Our analysis encompassed a comprehensive study of the existing data collection, consolidation, and reporting processes. We used flowcharts to capture and illustrate CLAARRDEC's project management activities. Furthermore, we actively sought their input regarding desired system features and functionalities. We also gathered pertinent data templates and sample reports currently managed by the office. From these insights, we conceptualized the primary components and activities for the proposed system (see Figure 1b).

We formulated top-level diagrams and technical design for the system: *i*) *wireframe diagram* to depict the fundamental structure and components of the system's user interface, along with navigation and key elements; *ii*) *use case* to delineate system functionality and its correlation with user types; *iii*) *flowcharts* for each primary function, offering a step-by-step visualization of the processes involved; and *iv*) *database design*.

## 3.2 Development and Testing

We utilized a server-side scripting language, PHP, coupled with the Laravel framework to serve as the backbone during the development phase. MySQL was employed as the database. For the frontend, we crafted the user interface using Bootstrap. Additionally, we integrated AdminLTE, a widely used open-source admin dashboard and control panel template. To implement the data visualization functions of the system, we leveraged JavaScript frameworks such as Vue.js and apexcharts.js.

We integrated a series of tests alongside coding activities. Modular testing sessions were conducted to rigorously test each feature of the system, exclusively within the development team. The modular tests identified errors related to database queries, form validation, missing fields, navigation issues, and unresponsive form elements. It is worth noting that utilizing Laravel in the development process underscores the crucial roles of controllers, models, and routes. Controllers manage incoming requests, models define data structure and behavior, and routes facilitate navigation. Modular testing ensured these components functioned correctly, addressing errors and inconsistencies iteratively until achieving a 100% pass rate. The modular tests established the reliability and functionality of the system's core elements.

Two levels of functional tests were performed to identify problems or bugs arising from the integrated Admin and User modules. Initially, the system underwent testing by IT experts to ensure correct behavior, intended functions, and expected outcomes. GUI elements, user authentication, report submission and management, input data processing, error handling, database operations, and the management of CMI, researcher profiles, and user accounts, were thoroughly tested. The functional integration tests by IT experts were conducted to validate the coherence of data and process flow between admin and user modules and ensure integrated features functioned correctly. Following refinements based on IT expert feedback, the second level of functional testing involved CMI focal persons (i.e., the users) and CLAARRDEC personnel (i.e., the admin). A training workshop was conducted, allowing the focal persons to input their actual data via the user modules, and the CLAARRDEC personnel to perform monitoring and report generation tasks via the admin modules.

After implementing further refinements based on feedbacks from the latter round of functional testing, two levels of acceptance tests were carried out to ensure the system met stakeholder expectations and was deployment-ready. The alpha test, conducted by the development team, included functional validation, performance testing, and compatibility testing. The beta test involved another workshop with CMI focal persons, executing prepared test cases with their real-world data to validate system readiness for deployment.

## 4 RESULTS AND DISCUSSION

### 4.1 The Database Model

Figure 2 illustrates the graph database model of the system, wherein entities and their relationships are depicted by nodes and edges, respectively. Numeric tags denote the number of properties for each entity. In this model, a 'program' denotes an R&D endeavor with extensive scope and prolonged duration, often comprising multiple interconnected projects serving a common purpose. A 'project' may either be a standalone initiative or part of a larger program, while more extensive projects may include smaller 'sub-projects'. The database tables, represented by entities, capture and store the data associated with every R&D undertaking of a Consortium Member Institution (CMI). This database design offers a robust framework for the system, facilitating effective management and organization of data pertaining to each R&D activity of CMIs. Moreover, the design is scalable, accommodating future enhancements and additional entities as necessary. The simplified structure and clear relationships ensure efficient querying and retrieval of information, supporting the generation of diverse reports required by CLAARRDEC.

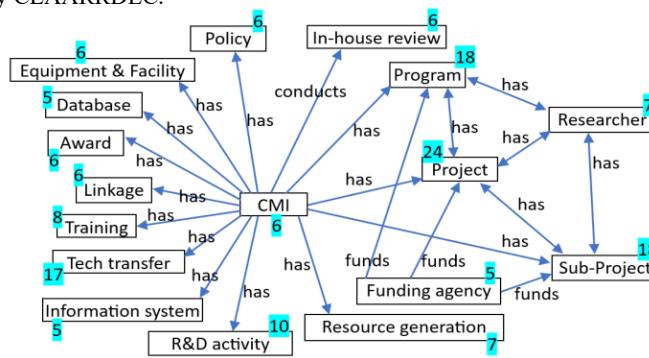

Figure 2: The data model of the system.

## 4.2 The System Features

Figures 3 shows the dashboard and key features accessible to admin users. It provides real-time visualizations of R&D programs/projects across CMIs. The features include creating new reports for programs, projects, and sub-projects, as well as viewing, editing, deleting reports, and managing user accounts, including password recovery. The CMI control panel offers a similar dashboard and features, albeit restricted to the CMI's specific data. A dedicated input form facilitates the entry of details for R&D engagements.

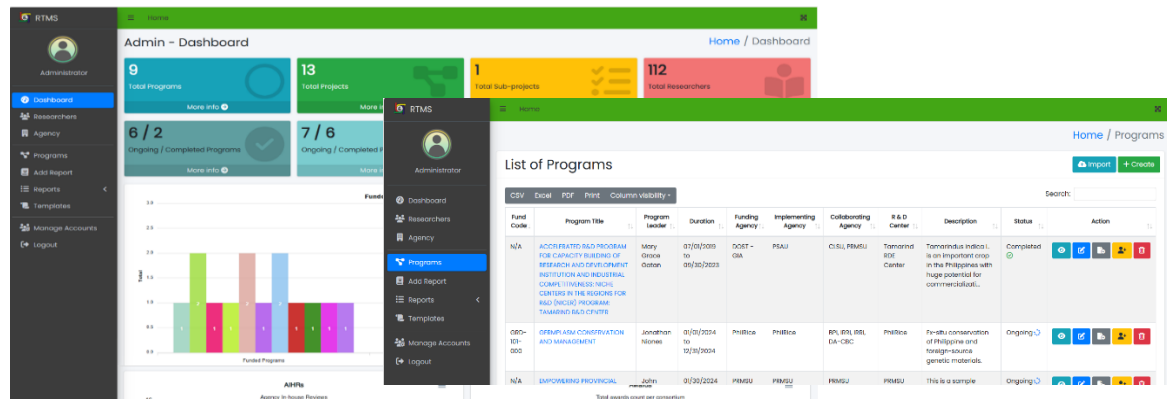

Figure 3: System features for the admin user.

The reports generation feature is another important component of the system, offering the capability to generate various types of reports beyond program and project reports. In total, there are 16 reports that can be submitted to the system using the 'Add Report' feature. These reports are categorized into five sections aligned with CLAARRDEC's Annual Report: R&D Management and Coordination, Strategic R&D Activities, R&D Results Utilization, Capability Building and Governance, and Policy Analysis and Advocacy. Figure 4 illustrates the current classification of reports generated by the system, along with the source database entities required for each report. These reports are consolidated into the annual report generated by CLAARRDEC. Additionally, the system allows for the generation of either full reports (i.e., annual report) or specific reports (i.e., filtered reports), providing flexibility in reporting options.

## 4.3 The Participatory Development Approach

It is noteworthy that we have employed a participatory development approach [6, 8] throughout our work. Stakeholders significantly contributed to all phases of our development process, from initial specification of needs and requirements to the final assessment of the system. For instance, during the design phase, wireframes and interface designs were shared with CLAARRDEC management and staff, as well as with the CMI focal persons. The feedback gathered from these interactions was instrumental in further refining the designs and ensuring alignment with user expectations. Moreover, our methodology has intertwined the development and testing phases, with iterative activities within each phase. The testing activities, which actively involved users, are documented, including detailed summaries of test cases, results, and descriptions of revisions and corresponding actions needed to address failed tests. For example, Table 1 shows that during one functional test, only 16 out of 28 test cases passed, indicating a 57% pass rate. Certain functionalities related to report submission, CMI management, and specific admin modules encountered failures, highlighting areas needing further refinement and debugging. This has resulted in substantial revisions, which encompassed adding new forms, adding or modifying form fields, revising name placeholders, incorporating form validation, and ensuring consistency across different sections of the system. This participatory approach has made our development more efficient, as potential revisions were not only identified during testing but also refined based on valuable feedback from participants.

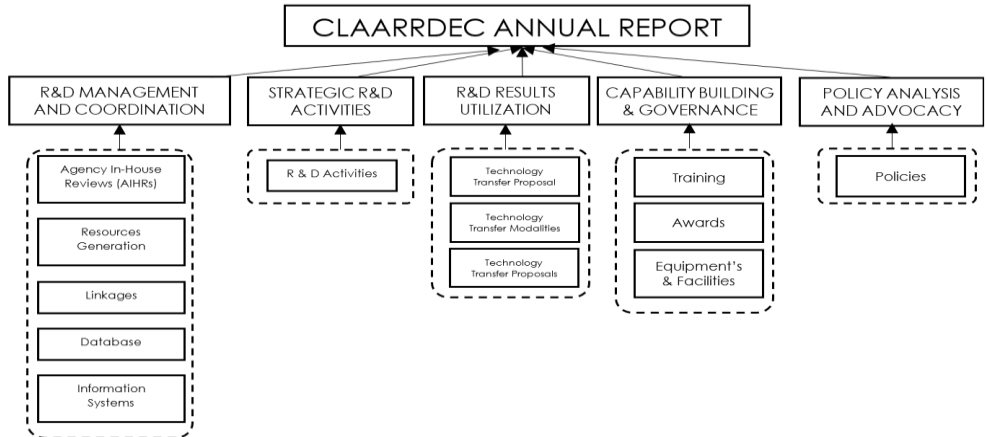

Figure 4: Categories of reports generated by the system.

The active participation of the intended users throughout the development process, especially their input for both the data acquisition and report functions of the system, has significantly contributed to the successful implementation. This participatory development approach, widely acknowledged in large-scale projects [8], serves to build trust among users and fosters a sense of their impact on the system's development. The approach has been documented in the literature (e.g., [6, 8]) to offer dual benefits. Firstly, *user empowerment* - involving users in the design process empowers them to take ownership of the final product, potentially leading to increased user satisfaction and adoption. End-users feel a sense of ownership, believing that their needs have been considered. Secondly, *alignment with user requirements and expectations* - through participatory design and prototyping, the final product is more likely to align with user expectations and preferences. This alignment is crucial for the success and acceptance of the system.

Table 1: Result of the various test phases.

| Test Phase | Test case pass rate | # of major feedback | # of revisions made |
|---|---|---|---|
| Modular/Unit Tests | 3/3 | - | - |
| Functional Test (L1-p1) | 16/28 (57%) | 21 | 26 |
| Functional Test (L1-p2) | 17/28 (61%) | 4 | 10 |
| Functional Test (L2) | 12/28 (43%) | 43 | 10 |
| Acceptance (Alpha) | 30/30 (100%) | 1 | 1 |

## 5  CONCLUSION AND FUTURE WORKS

The development of the web-based database management system and real-time monitoring solution represents a significant stride towards addressing the challenges faced by the Central Luzon Agriculture, Aquatic and Natural Resources Research and Development Consortium (CLAARRDEC). By leveraging digital technology, this system enhances the consortium's ability to effectively monitor and evaluate their R&D activities. The integration of features such as data collection, storage, retrieval, and utilization streamlines information sharing, promotes collaboration, and aids decision-making processes within CLAARRDEC. Moreover, the system empowers member institutions to establish and monitor their own R&D engagements, fostering a culture of accountability and innovation.

Looking ahead, we aim to explore the integration of blockchain and AI into our system (e.g., [1, 2]) to secure and automate project management processes. This advancement is anticipated to enhance the decision support capability of the system and foster its potential adoption across other research consortia throughout the Philippines.

ACKNOWLEDGMENTS

This work is partially supported by the Department of Science and Technology - Philippine Council for Agriculture, Aquatic and Natural Resources Research and Development of the Philippines.

REFERENCES

[1] Gunnar Auth, Oliver Jokisch, and Christian Durk. 2019. Revisiting automated project management in the digital age – a survey of AI approaches. Journal of Applied Knowledge Management, 7(1). doi: 10.36965/OJAKM.2019.7(1)27-39
[2] Yu Bai, Zehui Li, Kangning Wu, Jialiang Yang, Sheng Liang, Bocheng Ouyang, Zhiyu Chen, & Junyang Wang. 2018. Researchain: Union Blockchain Based Scientific Research Project Management System. In Proc. Chinese Automation Congress (CAC). doi: 10.1109/cac.2018.8623571
[3] A Boulle, A Heekes, N Tiffin, M Smith, T Mutemaringa, N Zinyakatira, F Phelanyane, C Pienaar, K Buddiga, E Coetzee, R Van Rooyen, R Dyers, N Fredericks, A Loff, L Shand, M Moodley, I De Vega, and K Vallabhjee. 2019. Data Centre Profile: The Provincial Health Data Centre of the western cape province, South Africa. International Journal of Population Data Science, 4(2). doi: 10.23889/ijpds.v4i2.1143
[4] Ningning Ge, Hui Li, Lingwang Gao, Zhiyuan Zang, Yi Li, Jie Li, Xihong Lei, and Zourui Shen. 2010. A web-based project management system for Agricultural Scientific Research. In Proc. International Conference on Management and Service Science. doi: 10.1109/icmss.2010.5576068
[5] Muhammad A. B. Ilyas, Mohammed Khalifa Hassan, and Muhammad Umara Ilyas. 2013. PMIS: boon or bane? In Proc. PMI® Global Congress EMEA, Istanbul, Turkey. [Online]. Available: https://www.pmi.org/learning/library/project-management-information-systems-overviews-5813
[6] Karlheinz Kautz. 2010. Participatory design activities and agile software development. In Proc. IFIP Working Conference on Human Benefit through the Diffusion of Information Design Science Research. doi: 10.1007/978-3-642-12113-5_18
[7] Li Li, Ping Li, Qing Liu, Jian Zhang, Zhongmin Wang, and Jungang Han. 2007. WebUPMS: a web-based undergraduate project management system. In Proc. IEEE International Symposium on Information Technologies and Applications in Education, Kunming, China. doi: 10.1109/ISITAE.2007.4409304
[8] Wendy E. Mackay and Michel Beaudouin-Lafon. 2023. Participatory design and prototyping. In: Vanderdonckt, J., Palanque, P., Winckler, M. (eds) Handbook of Human Computer Interaction, Springer, Cham, doi:10.1007/978-3-319-27648-9_31-1
[9] Rosa Micale, Concetta Manuela La Fata, Alberto Lombardo, and Giada La Scalia. 2021. Project management information systems (PMISs): a statistical-based analysis for the evaluation of software packages features. Applied Sciences 11(23). doi: 10.3390/app112311233
[10] Jean M. Moran, Mary Feng, Lisa A. Benedetti, Robin Marsh, Kent A. Griffith, Martha M. Matuszak, Michael Hess, Matthew McMullen, Jennifer Fisher, Teamour Nurushev, Margaret Grubb, Stephen Gardner, Daniel Nielsen, Reshma Jagsi, James A. Hayman, and Lori J. Pierce. 2017. Development of a model web-based system to support a statewide quality consortium in radiation oncology. Pract Radiat Oncol. doi: 10.1016/j.prro.2016.10.002.
[11] PMI. 2021. PMBOK® guide, seventh edition: a guide to the project management body of knowledge and standard for project management. Project Management Institute.
[12] Ignatius M. Prinsloo, Johan N. du Plessis, and Hendrik G. Brand. 2015. A data consolidation platform for a web-based energy information system. In Proc. IEEE International Conference on the Industrial and Commercial Use of Energy (ICUE), Cape Town, South Africa, doi: 10.1109/ICUE.2015.7280255
[13] Louis Raymond and Francois Bergeron. 2008. Project management information systems: an empirical study of their impact on project managers and project success. Intl Journal of Project Management, 26(2). doi: 10.1016/j.ijproman.2007.06.002
[14] Ines Silva, Diana Ferreira, Hugo Peixoto, and Jose Machado. 2023. A data acquisition and consolidation system based on openEHR applied to physical medicine and rehabilitation. Procedia Computer Science 220, 844-849. doi: 10.1016/j.procs.2023.03.113
[15] Zhang Yan, Guo Wei, Liu Dongdong, Niu Lei, and Yan Mengran. 2020. University research project management system based on cloud platform. In Proc. IEEE International Conference on Big Data and Informatization Education (ICBDIE), Zhangjiajie, China. doi: 10.1109/ICBDIE50010.2020.00112